\documentclass[12pt]{iopart}

\newcommand{\ul}[1]{\hspace{-0.1ex}\underline{\hspace{0.1ex} #1 \hspace{-0.1em}}\hspace{0.1em}}
\newcommand{\tauindicesstring}{underlining }

\newcommand{\bydef}{\equiv} 

\newcommand{\ts}{\textstyle}

\newcommand{\be}{\begin{equation}}
\newcommand{\ee}{\end{equation}}
\newcommand{\bea}{\begin{eqnarray}}
\newcommand{\eea}{\end{eqnarray}}
\newcommand{\bse}{\begin{subequations}}
\newcommand{\ese}{\end{subequations}}
\newcommand{\ifhat}{} 

\newcommand{\pp}{Q} 
\newcommand{\kk}{S} 
\newcommand{\hpp}[1]{{\ifhat \pp_{#1}}} 
\newcommand{\hkk}[1]{{\ifhat \kk^{#1}}} 

\newcommand{\opppp}[1]{{\left(#1\right)^{\alpha\beta} \{\hpp{\alpha}, \hpp{\beta}\} }} 
\newcommand{\okkkk}[1]{{\left(#1\right)_{\alpha\beta} \{\hkk{\alpha}, \hkk{\beta}\} }} 
\newcommand{\oppkk}[1]{{\left(#1\right)^\alpha_{\ \beta} \{\hpp{\alpha}, \hkk{\beta}\} }} 
\newcommand{\smopppp}[1]{{(#1)^{\alpha\beta} \{\hpp{\alpha}, \hpp{\beta}\} }}
\newcommand{\smokkkk}[1]{{(#1)_{\alpha\beta} \{\hkk{\alpha}, \hkk{\beta}\} }}
\newcommand{\smoppkk}[1]{{(#1)^\alpha_{\ \beta} \{\hpp{\alpha}, \hkk{\beta}\} }}

\newcommand{\fto}[1]{{\hat a_{#1}}} 
\newcommand{\ftoa}[1]{{\hat a^\dagger_{#1}}} 

\newcommand{\quadalg}{{\cal A}_2}

\newcommand{\DSx}{Y} 
\newcommand{\DS}[1]{Y_{\ul #1}} 

\usepackage{bm}

\begin{document}

\title{Parabose algebra as generalized conformal supersymmetry}

\author{Igor Salom}

\address{Institute of Physics,
11001 Belgrade, P.O. Box 57, Serbia}
\ead{isalom@phy.bg.ac.yu}
\begin{abstract}
The form of realistic space-time supersymmetry is fixed, by
Haag-Lopuszanski-Sohnius theorem, either to the familiar form of
Poincare supersymmetry or, in massless case, to that of conformal
supersymmetry. We question necessity for such strict restriction
in the context of theories with broken symmetries. In particular,
we consider parabose $N=4$ algebra as an extension of conformal
supersymmetry in four dimensions (coinciding with the, so called,
generalized conformal supersymmetry). We show that sacrificing of
manifest Lorentz covariance leads to interpretation of the
generalized conformal supersymmetry as symmetry that contains, on
equal footing, two "rotation" groups. It is possible to reduce
this large symmetry down to observable one by simply breaking one
of these two $SU(2)$ isomorphic groups down to its $U(1)$
subgroup.
\end{abstract}

\maketitle

\section{\label{sec1}Introduction}

Prospect of finding larger symmetries that would embed observable
Poincare symmetry and possibly some of the internal symmetries (in
a non trivial way) has attracted physicists for a long time. Early
attempts ended in formulation of the famous Coleman and Mandula
theorem \cite{CM}, but this no-go theorem was soon evaded by the
idea of supersymmetry. The Coleman and Mandula theorem was then
replaced by the Haag-Lopuszanski-Sohnius (HLS) theorem \cite{HLS}
that put the now standard super-Poincar\' e and super-conformal
symmetries at the place of maximal supersymmetries of realistic
models (up to multiplication by an internal symmetry group).
However, the attempts to go around these no-go theorems never
truly ceased, many of these trying to weaken the mathematical
requirements of the theorems \cite{ZaobilazenjaHLS1,
ZaobilazenjaHLS2}.

We will here consider an extension of conformal supersymmetry (and
thus also of the Poincar\' e supersymmetry) which does not meet
one of the physical premises of the no-go theorem \cite{CM} -
namely the "particle finiteness" premise: "for any finite M, there
are only finite number of particle types with mass less than M".
Regarding this requirement, S. Coleman \cite{Cpretecha} comments:
"We would probably be willing to accept a theory with an infinite
number of particles, as long as they were spread out in mass in
such a way that experiments conducted at limited energy could only
detect a finite number of them". Our point is that a proper
symmetry breaking can, in principle, induce such mass splitting.
Besides, this does not have to imply increasing of the complexity
of a theory, since symmetry breaking is already an inescapable
component of all supersymmetric models (and of the most of the
contemporary physical models in general).

In this paper we will demonstrate that (non-extended) conformal
superalgebra can be seen as a part of a very simple algebra,
namely of $N=4$ parabose algebra. Relation of this algebra with
the standard conformal superalgebra is rather interesting: it is
the algebra that is obtained from non-extended conformal
superalgebra when we remove the algebraic "constraints" $\{Q_\eta,
Q_\xi\} = 0$ (allowing these and adjoint anticommutators to be new
symmetry generators) and appropriately close the algebra. All
commutator relations, used to define the conformal superalgebra,
remain the same within this bigger algebra, as well as all nonzero
anticommutator relations (apart from value of one coefficient). It
is intriguing that enlarging the conformal superalgebra in this
way simplifies the algebra instead of complicating it, as the
structural relations of the larger symmetry are determined by only
two defining relations of parabose algebra. More importantly, we
will show that the symmetry breaking necessary to reduce this
symmetry to observable one is also very simple, both conceptually
and by its mathematical form.

However, it is obvious that number of particles in a
supermultiplet becomes infinite, since consecutive action of the
same supersymmetry generator in this case no longer annihilates a
state. Nevertheless, we argue that such an obvious disagreement
with experimental data is a problem of the qualitatively same type
as occurs in the standard Poincar\' e supersymmetry. Namely,
whereas in models with Poincar\' e supersymmetry we need a
symmetry breaking to induce mass differences between finite number
of super-partners, here the symmetry breaking should provide
ascending masses for infinite series of super-partners. We will
demonstrate existence of simple form of symmetry breaking that
reduces the symmetry down to the Poincar\' e group, altogether
with providing the mass splitting.

This type of generalization of Poincar\'e and conformal
superymmetry, obtained by allowing $\{Q_\eta, Q_\xi\}$
anticommutators to be noncentral, has been already investigated,
mostly in the context of branes and M theory
\cite{Azcarraga,Townsend,Bars,Townsend2,Ferrara,GauntlettEtAll,LukierskiToppan,LeePark}.
It has been pointed out that conclusions of HLS theorem are not
applicable to considerations of extended objects \cite{Townsend,
Bars, LeePark}. In this context the anticommutators that are
forbiden by HLS scheme (usually known as "tensorial central
charges") can be interpreted as charges carried by domain walls
\cite{Azcarraga, Bars, Ferrara, GauntlettEtAll, LukierskiToppan,
LeePark}. Most of the interest in these papers is focused on
higher dimensional cases, but generalized algebras in four
dimensions are also investigated \cite{GauntlettEtAll, BanLuk,
Fedoruk, BanLukSor, BanLukSor2}, in some cases even in context of
particles instead of branes \cite{BanLuk, BanLukSor, BanLukSor2,
ToppanKuznetsova}.

Our intention is to analyze the four dimensional case from purely
algebraic point of view, considering all the algebra operators as
generators of space-time symmetries. The idea is to explain the
excess of symmetry generators by pointing out a type of symmetry
breaking that could reconcile this large symmetry with
observation. By using approach based on parabose algebra (which is
determined essentially by only two defining relations) we would
like to emphasize mathematical simplicity of the generalized
conformal supersymmetry. We will show that sacrificing manifest
Lorentz covariance leads to interesting interpretation of the
generalized supersymmetry in four dimensions as symmetry of
spacetime possessing two "rotational" groups existing on equal
footing. Then we demonstrate that a breaking of one of these two
$SU(2)$ isomorphic groups down to its $U(1)$ subgroup can lead to
reduction of overall symmetry down to observable Poincar\'e
symmetry. By underlining the simplicity of generalized
supersymmetry and of the form of the required symmetry breaking,
we would like to point out that potential significance of this
symmetry is underestimated, in comparison to the attention given
to the standard Poincar\'e and conformal supersymmetry hypothesis.
As the generalized supersymmetry qualitatively differs from the
standard supersymmetry (for example, by predicting infinite
supermultiplets), existence of this alternative version of
supersymmetry should be kept on mind when it comes to interpreting
experimental data in future.

In order to establish connection of $N=4$ parabose algebra with
conformal superalgebra in 4 space-time dimensions, we will need to
introduce a special basis for expressing not only parabose
operators themselves, but also for expressing their
anticommutators. This basis, in which the connection becomes
manifest, will be introduced in the next section. In section
\ref{sec3} the (bosonic) conformal algebra is recognized as a
subalgebra of algebra of parabose anticommutators and the symmetry
breaking is considered. Supersymmetry generators will be included
in the picture in section \ref{sec4}. In the last section some
additional remarks will be given.

Throughout the text, Latin indices $i, j, k, \dots$ will take
values 1, 2 and 3, Greek indices from the beginning of alphabet
$\alpha, \beta, \dots$ will take values from 1 to 4 and will in
general denote Dirac-like spinor indices, $\eta$ and $\xi$ will be
two-dimensional Weyl spinor indices, while Greek indices from the
middle of alphabet $\mu, \nu, \dots$ will denote Lorentz
four-vector indices.

\section{\label{sec2}Special basis}

Parabose algebra with $N$ degrees of freedom \cite{Green} is
determined by trilinear relations %
\be
[\{ \fto{\alpha}, \fto{\beta}\}, \fto{\gamma}] = 0,  \qquad %
{}[\{ \fto{\alpha}, \ftoa{\beta}\}, \fto{\gamma}] =
-2\delta^\gamma_\beta \fto{\alpha},  \label{parabose algebra} \ee%
connecting $N$ operators $\fto{\alpha}$ and their hermitian
conjugates $\ftoa{\alpha}$ (curly brackets denote
anticommutator).\footnote{Relation $[\{ \fto{\alpha}, \fto{\beta}
\}, \ftoa{\gamma}] = 2\delta^\gamma_\beta \fto{\alpha} +
2\delta^\gamma_\alpha \fto{\beta}$ obtained from these two by
generalized Jacobi identities, as well as relations obtained from
these by hermitian conjugation, are also implied.} We will be
interested in case $N = 4$. Being a generalization of an algebra
of bosonic creation and annihilation operators, parabose algebra
is, along with parafermi algebra, usually considered in context of
parastatistics. This will not be the case in this paper.

In order to demonstrate connection of parabose algebra with
conformal supersymmetry, we will rewrite relations (\ref{parabose
algebra}) in a different, rather complicated basis of operators.

As a first step of this change of basis, we will switch from
operators $\fto{\alpha}$ and $\ftoa{\alpha}$ to their hermitian
combinations, defined as\footnote{Normalization is so chosen to
simplify later comparison with the standard form of supersymmetry
relations.}: %
\be \hkk{\alpha} \bydef (\fto{\alpha} + \ftoa{\alpha}), \qquad
\hpp{\alpha} \bydef -i (\fto{\alpha} - \ftoa{\alpha}).
\label{bose2Heis}\ee %

Relations (\ref{parabose algebra}) imply the following six
relations: %
\be \hspace{-2cm}
\renewcommand{\arraystretch}{1.4}
\begin{array}{ll} [\{
\hpp{\alpha}, \hpp{\beta}\}, \hpp{\gamma}] = 0, & [\{
\hkk{\alpha}, \hkk{\beta}\}, \hkk{\gamma}] = 0,\\ %
{} [\{ \hpp{\alpha}, \hpp{\beta} \}, \hkk{\gamma}] =
-4i\delta^\gamma_\beta \hpp{\alpha} - 4i\delta^\gamma_\alpha
\hpp{\beta},\qquad & [\{ \hkk{\alpha}, \hkk{\beta} \},
\hpp{\gamma}] = 4i\delta_\gamma^\beta \hkk{\alpha} +
4i\delta_\gamma^\alpha \hkk{\beta}, \\ %
{}[\{ \hkk{\alpha}, \hpp{\beta}\}, \hkk{\gamma}] =
4i\delta^\gamma_\beta \hkk{\alpha}, & [\{ \hpp{\alpha},
\hkk{\beta}\}, \hpp{\gamma}] = 4i\delta_\gamma^\beta \hpp{\alpha}.
  \end{array} \label{para-Heisenberg algebra} \ee %

These relations express commutators between basic anticommutators
$\{\hpp{\alpha}, \hpp{\beta}\}$, $\{\hpp{\alpha}, \hkk{\beta}\}$,
$\{\hkk{\alpha}, \hkk{\beta}\}$ and operators $\hkk{\alpha}$ and
$\hpp{\alpha}$ themselves. We can rewrite these relations using
some other basis of anticommutators, i.e.\ using some linear
combinations $A^{\alpha\beta}\{\hpp{\alpha}, \hpp{\beta}\}$,
$B^\alpha_{\ \beta}\{\hpp{\alpha}, \hkk{\beta}\}$ and
$C_{\alpha\beta}\{\hkk{\alpha}, \hkk{\beta}\}$, where matrices of
coefficients $A$, $B$ and $C$ take values from some basis set of 4
by 4 real matrices. For that purpose we introduce the following
matrix basis: we choose a set of six real matrices
$\sigma_i$ and $\tau_{\ul i}$, $i, \ul i = 1,2,3$ satisfying %
\be [\sigma_i, \sigma_j] = 2\:\! \varepsilon_{ijk} \sigma_k,\quad %
[\tau_{\ul i}, \tau_{\ul j}] = 2 \:\! \varepsilon_{\ul i \ul j \ul k} \tau_{\ul k},\quad %
[\sigma_i, \tau_{\ul j}] = 0, %
\label{sigma tau commutators} \ee %
as a basis of antisymmetric four by four real
matrices\footnote{One possible realization of such matrices is,
for example: $\sigma_1 = -i\sigma_y \times \sigma_x$, $\sigma_2 =
-i I_2 \times \sigma_y$, $\sigma_3 = -i\sigma_y \times \sigma_z$,
$\tau_1 = i\sigma_x \times \sigma_y$, $\tau_2 = -i \sigma_z \times
\sigma_y$, $\tau_3 = -i\sigma_y \times I_2$, where $\sigma_x$,
$\sigma_y$ and $\sigma_z$ are standard two dimensional Pauli
matrices and $I_2$ is a two dimensional unit matrix.}, and we
choose nine matrices $\alpha_{\ul ij} \bydef \tau_{\ul i}
\sigma_j$, plus the unit matrix denoted as $\alpha_0$, as a basis
for symmetric matrices. (Notice that we distinct tau indices from
sigma indices by \tauindicesstring the former.) Matrices $\tau$
and $\sigma$ are four dimensional analogs of Pauli matrices and
they are here defined to be anti-hermitian, satisfying $\sigma_i^2
= \tau_{\ul i}^2 = -1$.

Using these matrices we define new basis for expressing
anticommutators of $\hpp{}$ and $\hkk{}$: %
\be
  \renewcommand{\arraystretch}{1.5}
  \begin{array}{ll}
\hat J_i \bydef \frac 18 \oppkk{\displaystyle \sigma_i}, & \DS{i}
\bydef \frac 18 \oppkk{\displaystyle \tau_{\ul i}}, \\
\hat N_{\ul ij} \bydef \frac 18 \oppkk{\displaystyle \alpha_{\ul
ij}}, &
\hat D \bydef \oppkk{\displaystyle \alpha_0},  \\
\hat P_{\ul ij} \bydef \frac 18 \opppp{\displaystyle \alpha_{\ul
ij}},& \hat P_0 \bydef \frac 18 \opppp{\displaystyle \alpha_0},   \\
 \hat K_{\ul ij} \bydef -\frac 18 \okkkk{\displaystyle
\alpha_{\ul ij}}, & \hat K_0 \bydef \frac 18 \okkkk{\displaystyle
\alpha_0}.   \end{array} \label{Jeleb identification} \ee

By expressing relations (\ref{para-Heisenberg algebra}) in terms
of these linear combinations of anticommutators, we obtain: %
\be \hspace{-2cm} \renewcommand{\arraystretch}{1.4}
  \begin{array}{lll}
{}[\hat J_i, \hpp{\alpha}] = -i (\frac{\ts \sigma_i}{2})_\alpha^{\
\beta}\, \hpp{\beta}, & [\ifhat \DS{i}, \hpp{\alpha}] = -i
(\frac{\ts \tau_{\ul i}}{2})_\alpha^{\ \beta}\, \hpp{\beta}, &
[\hat N_{\ul ij}, \hpp{\alpha}] = i (\frac{\ts \alpha_{\ul
ij}}{2})_\alpha^{\ \beta}\, \hpp{\beta},\\%
{}[\hat J_i, \hkk{\alpha}] = -i (\frac{\ts \sigma_i}{2})^\alpha_{\
\beta}\, \hkk{\beta},& [\ifhat \DS{i}, \hkk{\alpha}] = -i
(\frac{\ts \tau_{\ul i}}{2})^\alpha_{\ \beta}\, \hkk{\beta}, &
[\hat N_{\ul ij}, \hkk{\alpha}] = -i
(\frac{\ts \alpha_{\ul ij}}{2})^\alpha_{\ \beta}\, \hkk{\beta},  \\ %
{}[\hat K_0, \hpp{\alpha}] = i (\alpha_0)_{\alpha\beta} \,
\hkk{\beta}, & [\hat K_{\ul ij}, \hpp{\alpha}] = -i (\alpha_{\ul
ij})_{\alpha\beta} \, \hkk{\beta}, & [\hat K_0, \hkk{\alpha}] =
[\hat K_{\ul ij}, \hkk{\alpha}] = 0,\\%
{}[\hat P_0, \hkk{\alpha}] = -i (\alpha_0)^{\alpha\beta} \,
\hpp{\beta}, & [\hat P_{\ul ij}, \hkk{\alpha}] = -i (\alpha_{\ul
ij})^{\alpha\beta} \, \hpp{\beta}, &
{} [\hat P_0, \hpp{\alpha}] = [\hat P_{\ul ij}, \hpp{\alpha}] = 0, \\%
{}[\hat D, \hpp{\alpha}] = i (\frac{1}{2}) \hpp{\alpha}, & [\hat
D, \hkk{\alpha}] = -i (\frac{1}{2}) \hkk{\alpha}.
\end{array} \label{bose-fermi commutators} \ee %
These relations, combined with definitions (\ref{bose2Heis}) and
(\ref{Jeleb identification}), are equivalent to two starting
relations of parabose algebra (\ref{parabose algebra}). The
extreme superficial complexity of these numerous relations stems
only from the complicated choice of variables, i.e.\ of basis
operators.

In the following sections we will clarify connection of relations
(\ref{bose-fermi commutators}) with conformal superalgebra.

\section{\label{sec3}Connection of $N=4$ parabose algebra with conformal algebra and the symmetry breaking}

It is not difficult to see that set of all anticommutators of
starting parabose operators forms an algebra, to be denoted as
$\quadalg$. It has 36 generators and is isomorphic to $sp(2n)$
algebra, where $n=4$. Operators defined by (\ref{Jeleb
identification}) represent a particular basis of this algebra. In
this basis structural relations of algebra $\quadalg$ have the
following form: %
\bea && [\ifhat J_i, \ifhat J_j] = i\, \varepsilon_{ijk} \ifhat
J_k,\quad [\ifhat \DSx_{\ul i}, \ifhat \DSx_{\ul j}] =  i\,
\varepsilon_{\ul i\ul j\ul k} \ifhat \DSx_{\ul k},
\quad [\ifhat J_i, \ifhat \DSx_{\ul j}] = 0, \nonumber \\
{}&& [\ifhat J_i, \ifhat N_{\ul jk}]=i \varepsilon_{ikl} \ifhat
N_{\ul jl}, \qquad [\ifhat \DSx_{\ul i}, \ifhat N_{\ul
jk}]=i \varepsilon_{ijl} \ifhat N_{\ul lk}, \nonumber \\
{} && [\ifhat N_{\ul ij}, \ifhat N_{\ul kl}] = -i
\left(\delta_{jl} \varepsilon_{\ul i \ul k \ul m} \ifhat\DSx_{\ul
m} + \delta_{\ul i \ul k} \varepsilon_{jlm}\ifhat
J_m\right),\nonumber
\\&& {}[\ifhat J_i, \ifhat D] = [\ifhat\DSx_{\ul
i}, \ifhat D] = [\ifhat N_{\ul ij}, \ifhat D] = 0,  \nonumber \\
&& [\ifhat J_i, \ifhat P_{\ul jk}]=i \varepsilon_{ikl} \ifhat
P_{\ul jl}, \qquad [\ifhat \DSx_{\ul i}, \ifhat P_{\ul jk}] =i
\label{bose commutators} \varepsilon_{\ul i \ul j \ul l} \ifhat
P_{\ul lk},
\\ {}&& [\ifhat N_{\ul ij}, \ifhat P_{\ul kl}]=i
\delta_{\ul i \ul k}\delta_{jl} \ifhat P_0 + i \varepsilon_{\ul
i\ul k\ul m}\varepsilon_{jln} \ifhat P_{\ul mn} ,  \nonumber \\
{}&& [\ifhat N_{\ul ij}, \ifhat P_0]=i \ifhat P_{\ul ij}, \qquad
[\ifhat D, \ifhat P_{\ul ij}]=i \ifhat P_{\ul ij}, \nonumber \\ &&
{} [\ifhat D, \ifhat P_0]=i \ifhat P_0, \qquad [\ifhat J_i, \ifhat
P_0] = [\ifhat \DSx_{\ul i}, \ifhat P_0]=0, \nonumber \\ &&
[\ifhat J_i, \ifhat K_{\ul jk}]=i \varepsilon_{ikl} \ifhat K_{\ul
jl}, \qquad [\ifhat \DSx_{\ul i}, \ifhat K_{\ul jk}] =i
\varepsilon_{\ul i \ul j \ul l} \ifhat K_{\ul lk}, \qquad \dots
\nonumber \\&& [\ifhat P_{\ul ij}, \ifhat K_{\ul kl}]= 2i \left(
\delta_{\ul i\ul k}\delta_{jl} \ifhat D + \varepsilon_{\ul i\ul
k\ul m}\varepsilon_{jln} \ifhat N_{\ul mn} - \delta_{\ul i\ul
k}\varepsilon_{jlm}\ifhat J_m - \delta_{jl}\varepsilon_{\ul i\ul
k\ul m}\ifhat \DSx_{\ul m}\right), \nonumber \\ && {} [\ifhat
P_{\ul ij}, \ifhat K_0]=2i \ifhat N_{\ul
ij}, \qquad [\ifhat P_0, \ifhat K_{\ul ij}]=2i \ifhat N_{\ul ij}, \nonumber \\
{} && [\ifhat P_0, \ifhat K_0]= 2i \ifhat D.  \nonumber\eea

We will now show that algebra $\quadalg$ has conformal algebra as
a subalgebra, and that the reduction from the corresponding group
to the conformal subgroup can be seen as a consequence of symmetry
breaking of one $SU(2)$ group to its $U(1)$ subgroup.

To obtain conformal subalgebra let us discard all operators from
$\quadalg$ basis (\ref{Jeleb identification}) with underlined
index having values $\ul 1$ and $\ul 2$. What we are left with is
a subalgebra isomorphic with conformal algebra $c(1,3)$ plus one
additional generator that commutes with the rest of the
subalgebra. The remaining operators that generate $c(1,3)$ algebra are: %
\be J_k, N_i \bydef N_{\ul 3 i}, D,
  P_i \bydef P_{\ul 3 i}, P_0, K_i \bydef K_{\ul 3
i}, K_0, \label{conformal identification} \ee %
playing roles of rotation generators, boost generators, dilatation
generator, momenta and pure conformal generators, respectively.
The additional remaining operator is $\DS{3}$ which commutes with
all of the conformal generators.

Alternatively, we could have obtained conformal subalgebra by
keeping operators with underlined index equal to $\ul 1$ or $\ul
2$, instead of $\ul 3$. As the matter in fact, if we pick any
linear combination of operators $\DS{i}$, or of operators $J_i$,
the subalgebra of $\quadalg$ that commutes with the chosen
operator will be $c(1,3) $ isomorphic. On the other hand,
operators $\ifhat \DS{i}$ and $\ifhat J_i$ constitute two,
mutually commuting $su(2)$ isomorphic subalgebras (a consequence
of $so(4) = su(2) \oplus su(2)$ identity). The two corresponding
$SU(2)$ isomorphic groups act, respectively, on underlined and on
non-underlined indices of algebra operators. Furthermore, if we
consider the way in which the recognized conformal subalgebra fits
into the larger algebra $\quadalg$, we see that spatial momenta,
being equal to $\frac 18 \smopppp{ \alpha_{\ul 3 j}}$ naturally
fit into a set of nine operators $\frac 18 \smopppp{ \alpha_{\ul
ij}}$, spatial components of pure conformal generators fit into a
set of nine $-\frac 18 \smokkkk{ \alpha_{\ul ij}}$ and boosts into
set of nine $\frac 18 \smoppkk{ \alpha_{\ul ij}}$. Overall, the
situation slightly looks like as if we had two independent
rotation groups, generated by $\DS{i}$ and $J_i$, while "momenta",
"boosts" and "pure conformal operators" were here determined by
two independent three-vector directions, each related to its own
"rotation" group. And the symmetry reduction from $\quadalg$ group
to its conformally isomorphic subgroup can be therefore understood
as a consequence of symmetry breaking of one of these two $SU(2)$
subgroups. Without loss of generality, we have assumed breaking of
the group generated by $\ifhat \DS{i}$, with $\ifhat \DS{3}$
generating the remaining $U(1)$ symmetry.

As a more concrete example of such symmetry breaking, we can
assume existence of effective potential being an increasing
function of absolute value of $\DS{3}$ [e.g.\ proportional to the
$(\DS{3})^2$]. If the potential is sufficiently strong, all low
energy physics would be constrained to subspace of $\DS{3}$
eigenvalue equal to zero, and the remaining symmetry would be
conformal symmetry. Moreover, since such a potential would have to
break dilatational symmetry, overall symmetry would be reduced to
the observable Poincar\' e group. (This is just the simplest of
all possibilieties. For example, the minimum of potential does not
have to be at $\DS{3} = 0$, or potential could be function of some
additional observables combined with $\DS{3}$. As long as the
potential changes with the change of $\DS{3}$, the extra
generators $P_{\ul 1i},P_{\ul 2i},K_{\ul 1i},K_{\ul 2i},N_{\ul
1i},N_{\ul 2i},\DS{1},\DS{2}$ will be broken. After all, a
preferred direction with respect of the $\DSx$ rotation group
could be introduced in some completely different way.)

Note that such symmetry breaking also automatically fixes metric
of space-time (i.e.\ of the remained symmetry) to be Minkowskian
[the remaining $J_i$ and $N_{\ul 3i}$ hermitian operators generate
exactly an $so(1,3)$ algebra]. It is also interesting that the
energy operator $P_0$ singles out among other momentum operators
(i.e.\ among the rest of operators quadratic in $\hpp{}$) even
before the symmetry reduction. Indeed, this operator, being the
sum of squares of $\hpp{\alpha}$, stands out as a positive
operator, and there is no algebra automorphism that takes any
other "momentum component" $P_{\ul ij}$ into the $P_0$ or
vice-versus. This gives us some right to interpret the full group
generated by $\quadalg$ as a symmetry that differs from the
observable space-time symmetry in the first place by existence of
two "spatial-like" rotations, whereas it possesses something that
looks like unique role of one axis (to be interpreted as the time
axis). And the symmetry breaking only gets us rid of one of the
"rotation-like" groups.

\section{\label{sec4}Supersymmetry generators}

Next, we turn attention to the role of operators $\hpp{}$ and
$\hkk{}$. First we note that under the action of generators from
conformal subalgebra (\ref{conformal identification}), they
transform exactly as supersymmetry generators in the standard
conformal superalgebra. To see this more clearly, we can introduce
the following (Majorana) representation of Dirac matrices: \be
\gamma_0 = i\tau_{\ul 2},\quad \gamma_i = \gamma_0 \;\!
\alpha_{\ul 3i} = i\tau_{\ul 1}\sigma_i, \quad \gamma_5 = -i
\gamma_0 \gamma_1 \gamma_2
\gamma_3 = i \tau_{\ul 3}. \label{alpha and gamma connection}\ee %
Relations (\ref{bose-fermi commutators}) [more precisely, that
part of these relations with conformal subalgebra operators
(\ref{conformal identification})], expressed by using these
matrices, gain the form familiar from the standard conformal
superalgebra. For example, commutators of Lorentz subalgebra
generators with parabose operators $\hpp{\alpha}$ can be now
written in the standard form $[\ifhat M_{\mu\nu}, \hpp{\alpha}] =
-i({\textstyle\frac 14} [\gamma_\mu, \gamma_\nu])_\alpha^{\ \beta}
\hpp{\beta}$ (where $\ifhat M_{ij}=\varepsilon_{ijk} \ifhat J_k$,
$\ifhat M_{i0} = \ifhat N_{\ul 3i} $), and so on. In particular,
we conclude that hermitian operators $\hpp{}$ and $\hkk{}$ are
Majorana spinors, written in Majorana basis of Dirac matrices
(\ref{alpha and gamma connection}).

Relations (\ref{Jeleb identification}) defining our "basis", used
for expressing anticommutators of parabose operators, can be
inverted to yield the following identities: %
\be  \renewcommand{\arraystretch}{1.5}
 \begin{array}{lr}
\{\hpp{\alpha},\hpp{\beta}\} = (\alpha_0)_{\alpha\beta}\, \ifhat
P_0 + (\alpha_{\ul ij})_{\alpha\beta}\, \ifhat P_{\ul
ij}, \\%
\{\hkk{\alpha},\hkk{\beta}\} = (\alpha_0)^{\alpha\beta}\, \ifhat
K_0 - (\alpha_{\ul ij})^{\alpha\beta}\, \ifhat K_{\ul ij},\\ %
\{\hkk{\alpha},\hpp{\beta}\} = (\alpha_0)^\alpha_{\ \beta}\,
\ifhat D + (\alpha_{\ul ij})^\alpha_{\ \beta}\, \ifhat N_{\ul ij}
+ (\sigma_{i})^\alpha_{\ \beta}\, \ifhat J_{i} + (\tau_{\ul
i})^\alpha_{\ \beta}\, \ifhat \DS{i}.
\end{array} \label{fermi anticommutators} \ee

These relations can be compared, using representation of Dirac
matrices (\ref{alpha and gamma connection}), to the anticommutator
relations of the standard conformal superalgebra. An obvious
difference is appearance of additional bosonic generators $P_{\ul
1i},P_{\ul 2i},K_{\ul 1i},K_{\ul 2i},N_{\ul 1i},N_{\ul
2i},\DS{1},\DS{2}$ in (\ref{fermi anticommutators}), which do not
exist in the standard conformal superalgebra (these operators,
i.e.\ these linear combinations of $\hpp{}$ and $\hkk{}$
anticommutators are, in the standard superalgebra, defined to be
zero). However, apart from this, it turns out that the only
difference is in the coefficient multiplying operator $\DS{3}$.
Namely, by comparing the commutation relations of the two
algebras, we recognize that operator $2\DS{3}$ plays the role of
chiral $R$-charge. (It is interesting that, in this picture,
chiral $R$-charge becomes part of an $su(2)$ subalgebra. This
subalgebra appears in unbroken symmetry on the same footing as the
rotational subalgebra.) The corresponding $R$ coefficient in the
case of conformal superalgebra has a different value that equals
3, fixed there only by the generalized Jacoby identities.

To summarize this comparison, the transition from $N=4$ parabose
algebra to non-extended conformal superalgebra is achieved by
setting $P_{\ul 1i} = P_{\ul 2i} = K_{\ul 1i} = K_{\ul 2i} =
N_{\ul 1i} = N_{\ul 2i} = \DS{1} = \DS{2} = 0$ and by replacing
the value of coefficient multiplying $\DS{3}$ operator.

The connection in the opposite direction (from conformal
superalgebra to this extension) can be established if we notice
that anticommutator of two left-handed $\hpp{}$ operators
$\{\hpp{\eta}, \hpp{\xi}\}$, or of two right-handed operators
$\{\overline Q_{\dot \eta}, \overline Q_{\dot \xi}\}$, yields
linear combination of operators $P_{\ul 1i}$ and $P_{\ul 2i}$ (and
similarly, such anticommutators of $\hkk{}$ operators yield
combinations of $K_{\ul 1i}$ and $K_{\ul 2i}$). Graded algebra
consisting of parabose operators and their anticommutators
[isomorphic\footnote{For general and more formal treatment of
connection between parabose algebras with Lie (super)algebras see,
for example, \cite{OSPVeza, komentarA1}.} to $osp(1,8)$] can be
seen as a special non-extended conformal superalgebra where all
anticommutators of supersymmetry generators are allowed to be
nonzero operators (so called "generalized conformal algebra"
\cite{LukierskiToppan, BanLukSor}).

As already announced in the introduction, by relaxing the
"constraint" $\{\hpp{\eta}, \hpp{\xi}\} = 0$, supermultiplets
become infinite. Nevertheless, the simple symmetry breaking
assumption, discussed in the previous section, breaks not only
extra bosonic generators, but also the supersymmetry generators
$\hpp{\alpha}$ and $\hkk{\alpha}$. Since action of operators
$\hpp{\eta}$ and $\hpp{\dot \eta}$ change value of $\DS{3}$ for
$\frac 12$, each following member of a supermultiplet would gain
higher and higher mass, whereas the low-energy space-time symmetry
would be given by the Poincar\' e group.

\section{\label{sec5}Conclusion}

In this paper we analyzed generalized conformal supersymmetry in
$D=4$ from algebraic point of view, constructing it using parabose
operators. By considering the way the conformal subalgebra fits
into the parabose algebra, we offered interpretation that the
whole symmetry should correspond to a space-time with two
"rotational groups" existing {\it a priori} on equal footing, of
which one should be broken in order to obtain observable symmetry.
This aspect of generalized supersymmetry is invisible unless we
sacrifice manifest Lorentz covariance. For example, the existence
of two "rotation" groups generating algebra automorphisms is
obscured if we write the first of relations (\ref{fermi
anticommutators}) in a more standard Lorentz covariant way
\cite{Townsend, Bars, Ferrara, BanLuk}:%
\be \{Q_\alpha, Q_\beta\} = (C\gamma^\mu)_{\alpha\beta}P_\mu +
(C\gamma^{\mu\nu})_{\alpha\beta}Z_{\mu\nu}, \ee%
with Lorentz antisymmetric tensor $Z_{\mu\nu}$ denoting the
components of generalized momentum other than four-momentum.
Notice that $P_\mu$ and $Z_{\mu\nu}$, which were in our case
connected by $\DSx$ rotations, even have different number of
Lorentz indices. The emergence of Minkowskian metric is, in this
picture, also a consequence of the symmetry breaking (the metric
need not be introduced by hand).

It is interesting that, although the analyzed symmetry is higher
and mathematical structure thus richer, the algebra relations are
actually simplified. Namely, commutators of bosonic with fermionic
operators (\ref{bose-fermi commutators}) are nothing more than
simple relations of parabose algebra written in a complicated
basis. Moreover, the fermionic anticommutators (\ref{fermi
anticommutators}) are relations that describe this new basis, so
these relations can be seen as a specific naming convention for
linear combinations of $\hpp{}$ and $\hkk{}$ anticommutators. The
idea is that this complicated basis becomes physically relevant
due to the symmetry breaking, analyzed in section \ref{sec3}. The
relatively simple symmetry breaking is therefore responsible not
only for reduction of the starting symmetry and for introduction
of mass splitting, but also for superficial complexity that hides
simplicity of the starting parabose algebra. Bosonic algebra
$\quadalg$ relations (\ref{bose commutators}) are direct
consequence of (\ref{fermi anticommutators}) and (\ref{bose-fermi
commutators}).

From the perspective of this higher symmetry, those relations of
standard conformal superalgebra that set some of the
anticommutators to zero appear as a kind of artificial constraints
-- constraints that are, in this picture, consequences of a
symmetry breaking. This fact, that some linear combinations of
anticommutators are zero (in standard superalgebra) makes it
impossible to see anticommutators of fermionic generators simply
as a naming convention, as it was possible for (\ref{fermi
anticommutators}).

Transition from the non-extended standard conformal superalgebra
to the symmetry discussed here can be done by allowing all
anticommutators of supersymmetry generators to be nonzero
operators. By doing so we end up with an algebra determined by
only two parabose relations (\ref{parabose algebra}).

We remind that at the present point supersymmetry is still only a
theoretical construct still awaiting for experimental
confirmation, and that, in particular, we possess no experimental
data that would suggest that supersymmetry, if exists, should be
of the form of the standard Poincar\'e (or conformal) type. Thus,
by putting forward the simplicity of the generalized supersymmetry
algebra in parabose formulation, as well as by demonstrating
simplicity of the required form of symmetry breaking, we would
like to point out that generalized supersymmetry should be
seriously considered as possible candidate for real space-time
supesymmetry, together with the conventional supersymmetry obeying
the HLS conclusions.



\section*{References}

\begin{thebibliography}{17}


\bibitem{CM} Coleman S and Mandula J 1967 {\it Phys. Rev.} {\bf 159} 1251

\bibitem{HLS} Haag R, Lopuszanski J T and Sohnius M F 1975 {\it Nucl. Phys.} {\bf B88}
257

\bibitem{ZaobilazenjaHLS1} Rausch de Traubenberg M and
Slupinski M J, 2002 {\it J. Math. Phys.} {\bf 43} 5145

\bibitem{ZaobilazenjaHLS2} Dam H and Biedenharn L 1979 {\it Phys. Lett. } B {\bf 81}
313

\bibitem{Cpretecha} Coleman S 1965 {\it Phys. Rev.} {\bf 138} 1262

\bibitem{Azcarraga} Azc\'arraga J A, Gauntlett J P, Izquierdo J M and
Townsend P K 1989 {\it Phys. Rev. Lett.} {\bf 63} 2443

\bibitem{Townsend} Townsend P K 1995 {\it Proc. of
PASCOS/Hopkins} ({\it Preprint} hep-th/9507048)

\bibitem{Bars} Bars I 1996 {\it Phys. Rev. } D {\bf 54} 5203 ({\it Preprint}
hep-th/9604139)

\bibitem{Townsend2} Townsend P 1997 {\it Cargese Lectures} ({\it Preprint} hep-th/9712004)

\bibitem{Ferrara} Ferrara S and Porrati M 1998 {\it Phys. Lett. } B {\bf 423} 255 ({\it Preprint} hep-th/9711116).

\bibitem{GauntlettEtAll}  Gauntlett J P, Gibbons G W, Hull C M and Townsend P K
2001 {\it Commun. Math. Phys.} {\bf 216} 431

\bibitem{LukierskiToppan}  Lukierski J and Toppan F 2002 {\it Phys. Lett. } B {\bf 539} 266

\bibitem{LeePark} Lee S and Park J H 2004 {\it J. High Energy Phys.} {\bf 06} 038

\bibitem{BanLuk} Bandos I and Lukierski J 1999 {\it Mod. Phys. Lett.} A {\bf 14} 1257

\bibitem{Fedoruk} Fedoruk S and Zima V G 2000 {\it Mod. Phys. Lett.} A  {\bf 15} 2281

\bibitem{BanLukSor} Bandos I, Lukierski J and Sorokin D 1999
Generalized Superconformal Symmetries and Supertwistor Dynamics
{\it Preprint} hep-th/9912051v1;

\bibitem{BanLukSor2} Bandos I, Lukierski J and Sorokin D 2000 {\it Phys.Rev.} D {\bf 61} 045002

\bibitem{ToppanKuznetsova} Kuznetsova Z and Toppan F 2005 {\it J. High Energy Phys.} {\bf 0505} 060

\bibitem{Green} Green H S 1952 {\it Phys. Rev.} {\bf 90} 270

\bibitem{OSPVeza} Palev T D 1980 {\it J. Math. Phys.} {\bf 21} 797

\bibitem{komentarA1} Plyushchay M S 1997 {\it Nucl. Phys.} B {\bf 491} 619.

\end{thebibliography}
\end{document}